\documentclass{PoS}

\newcommand*{\no}{\noindent}
\newcommand*{\bea}{\begin{eqnarray}}
\newcommand*{\eea}{\end{eqnarray}}

\newcommand*{\pref}[1]{(\ref{#1})}

\newcommand*{\nn}{\nonumber}

\newcommand*{\bma}{\begin{matrix}}
\newcommand*{\ema}{\end{matrix}}

\newcommand*{\la}{\langle}
\newcommand*{\ra}{\rangle}

\title{Probing standard-model Higgs substructures using tops and weak gauge bosons}

\ShortTitle{Higgs substructure}

\author{\speaker{Axel Maas}$^{1,+}$, Simon Fernbach$^1$, Lukas Lechner$^2$, Simon Pl\"atzer$^3$\thanks{Supported in part by the European Union's Horizon 2020 research and innovation program as part of the Marie Sk\l{}odowska-Curie Innovative Training Network MCnetITN3 (grant agreement no. 722104) and partial support by the COST actions CA16201 ``PARTICLEFACE'' and CA16108 ``VBSCAN''.}, Robert Sch\"ofbeck$^2$, Pascal T\"orek$^1$\\
        $^1$Institute of Physics, NAWI Graz, University of Graz, Universit\"atsplatz 5, 8010 Graz, Austria\\
        $^2$Institute for High Energy Physics, Nikolsdorfer Gasse 18, 1050 Wien, Austria\\
        $^3$Particle Physics, Faculty of Physics, University of Vienna, 1090 Wien, Austria\footnote{University of Vienna preprint number UWTHPH-2019-33.}\\
        $^+$E-mail: \email{axel.maas@uni.graz.at}}

\abstract{Manifest gauge-invariance requires that observable states in the standard-model are described by composite operators, which involve additional Higgs contributions beyond perturbation theory. This field-theoretical effect has been confirmed in lattice simulations. It should also be experimentally accessible at high enough precision. Here a few estimates for such signatures at current and future collider experiments will be discussed.}

\FullConference{
European Physical Society Conference on High Energy Physics - EPS-HEP2019 -\\
			10-17 July, 2019\\
			Ghent, Belgium}

\begin{document}

\section{Introduction}

The electroweak sector of the standard model appears at first sight to be straightforward, due to its weak coupling. Especially, its elementary degrees of freedom appear to be a suitable description of the observed particle spectrum at LHC \cite{pdg}. However, this seems less obvious when viewing it as a non-Abelian gauge theory. In such a theory, elementary degrees of freedom cannot be observable particles, not even as intermediate resonances, no matter how small the coupling \cite{Frohlich:1980gj,Frohlich:1981yi}. This arises mainly because of the geometric structure of non-Abelian groups. Hence, only composite objects should be observable.

This apparent contradiction is resolved by the Fr\"ohlich-Morchio-Strocchi (FMS) mechanism and the particular structure of the standard model \cite{Frohlich:1980gj,Frohlich:1981yi}. As will be detailed below, this lets the composite objects behave almost like the elementary ones. But only almost, and there are corrections, which are typically suppressed by powers of the ratio of the energy scale to the scale of the theory as set by the Higgs vacuum expectation value. This theoretical picture has been supported by lattice simulations. For a review see \cite{Maas:2017wzi}. There are now two very interesting consequences.

One is that theories, which are not as special as the standard model, can potentially have large, even qualitative, deviations between the elementary particle spectrum and the observable composite states, even at very weak coupling \cite{Maas:2015gma,Maas:2017xzh,Sondenheimer:gut}. This has also been confirmed in lattice simulations \cite{Maas:2018xxu}, see again \cite{Maas:2017wzi} for a review. The implications of this can substantially alter which theories are consistent extensions of the standard model \cite{Maas:2015gma,Maas:2017wzi,Maas:2017xzh,Sondenheimer:gut}.

However, in absence of new physics, as reported at this conference, such predictions are hard to test experimentally. Here the second consequence comes into play. If the energy scale is high enough, or the sensitivity is good enough, also in the standard model deviations can occur. Thus, in principle, this field-theoretical effect is accessible in experiment. In the following, a number of results will be reported \cite{Egger:2017tkd,Maas:2018ska,Fernbach:unpublished,Sondenheimer:unpublished}, in which it is attempted to estimate the size of the effects.

\section{Origin of the effect}

The technically simplest example to understand the origin of the additional contributions is the scalar particle observed. It is usually identified with the quantum fluctuation $\eta$ of the elementary Higgs field $\phi=v+\eta+\omega$ of the standard model around its vacuum expectation value $v$ and without its Goldstone components $\omega$. However, this field is not gauge-invariant\footnote{Note that the 'breaking' of the electroweak gauge symmetry is only a figure of speech, and in a field-theoretical clean setting it is never broken, see \cite{Maas:2017wzi} for a review. That this slang actually describes the experimental situation quite nicely at first sight is again a consequence of the FMS mechanism.}, and thus cannot describe an observable object \cite{Frohlich:1980gj,Frohlich:1981yi,Maas:2017wzi}. Rather, it is necessary to consider gauge-invariant object, like the composite operator $(\phi^\dagger\phi)(x)$. Such an operator has the same structure as, e.\ g., a meson operator in QCD, and describes a scalar bound state.

The FMS mechanism is now a straightforward decomposition of this bound state in a technically suitable gauge. In the 't Hooft gauge, the connected part of the propagator is given by
\bea
\la (\phi^\dagger\phi)(x)^\dagger (\phi^\dagger\phi)(y)\ra=4v^2\la \eta(x)\eta(y)\ra+v\la\eta(x)\eta(x)\eta(y)+\eta(y)\eta(y)\eta(x)\ra+\la\eta(x)\eta(x)\eta(y)\eta(y)\ra\nn\\\label{higgs}.
\eea
\no The leading term in $v$ is just the ordinary elementary Higgs propagator: At leading order in $v$ the composite propagator is identical to the propagator of the elementary Higgs. Especially, this implies that both have the same mass and width. Because $v$ is a large number compared to the quantum fluctuations of the Higgs field, this explains why so far the lower-order terms were not experimentally relevant. A similar expansion holds for any other observed state in the standard-model \cite{Frohlich:1980gj,Frohlich:1981yi,Maas:2017wzi}. However, degeneracies now become linked to either global symmetries or QED. E.\ g., the triplet of $Z$ and $W^\pm$ are no longer a triplet, up to mixing, of the weak isospin, but of the custodial symmetry. Likewise, intrageneration flavor of left-handed fermions becomes replaced by custodial symmetry \cite{Frohlich:1980gj,Frohlich:1981yi}, which is also true for open-flavor hadrons \cite{Egger:2017tkd}. In quantum-number channels without elementary counter part only scattering states of composite particles are found, yielding a complete identification of elementary states and composite states \cite{Maas:2017wzi}. This perfect mapping is because of the interplay of the weak isospin group and the custodial group, and because both are SU(2).

But there are the further terms in \pref{higgs}. Even if there are no non-perturbative effects in the further terms, they contribute perturbatively. E.\ g., to all orders in $v$, but at tree-level in all other couplings, the bound state propagator reads in the pole scheme \cite{Maas:2017wzi,Sondenheimer:unpublished}
\bea
D(p)&=&\la (\phi^\dagger\phi)^\dagger (\phi^\dagger\phi)\ra(p)\approx\frac{1}{p^2-m_H^2+4v^2 \Pi(p)+i\epsilon}\nn\\
\Pi(p)&=&-\frac{i\pi^2}{2}\left(\frac{m_H^2}{p^2}\left(\frac{1}{r}-r\right)\ln r\right)\nn\\
r&=&\frac{-p^2+2m_H^2-i\epsilon\pm\sqrt{(p^2-2m_H^2+i\epsilon)^2-4m_H^4}}{2m_H^2}\label{sprop},
\eea
\no where $m_H$ is the observed mass of the scalar and identified with the Higgs mass.

\begin{figure}
\includegraphics[width=\linewidth]{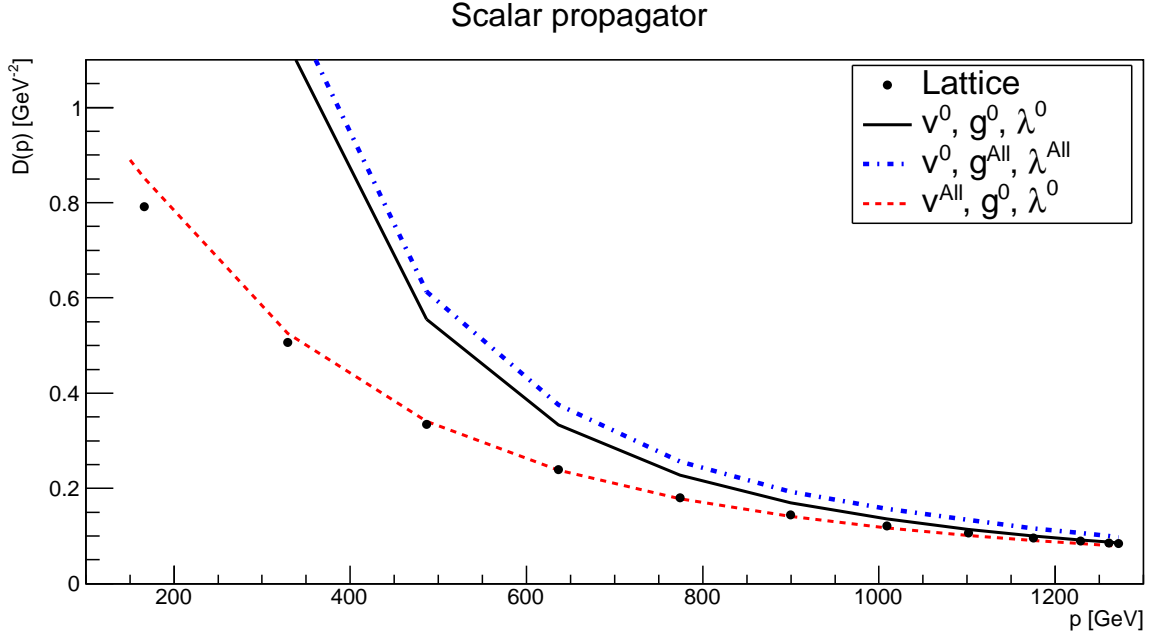}
\caption{\label{pt}The propagator of the composite scalar from the lattice at spacelike momenta. The result is compared to the tree-level elementary Higgs propagator, the all-order result for the elementary Higgs propagator obtained from lattice results \cite{Maas:2013aia}, and a comparison to the FMS result \pref{sprop}. In the conventions of \cite{Maas:2017wzi} the parameters are gauge coupling $g=1.46$, 4-Higgs coupling $\lambda=2.14$, and Higgs vacuum expectation value $v=55$ GeV.}
\end{figure}

This prediction is compared to lattice results in figure \ref{pt}. As is visible, inclusion of higher orders in the vacuum expectation value yields much better agreement than even inclusion of all orders in the other coupling parameters but using only the leading order in the vacuum expectation value in \pref{higgs}. Including higher orders in this augmented perturbation theory \cite{Maas:2017wzi} would yield even better agreement, as non-perturbative contributions in the correlation functions on the right-hand side of \pref{higgs} appear to be negligible \cite{Maas:2013aia}. Thus, the correlation function of a composite operator can be determined perturbatively if augmented by the FMS mechanism.

\section{Phenomenological consequences}

The propagator is, of course, not a directly measurable quantity. However, the same augmented perturbation theory should also be valid when calculating observables, like cross sections \cite{Egger:2017tkd,Maas:2017wzi}. This will be discussed now for two examples.

\begin{figure}
\includegraphics[width=\linewidth]{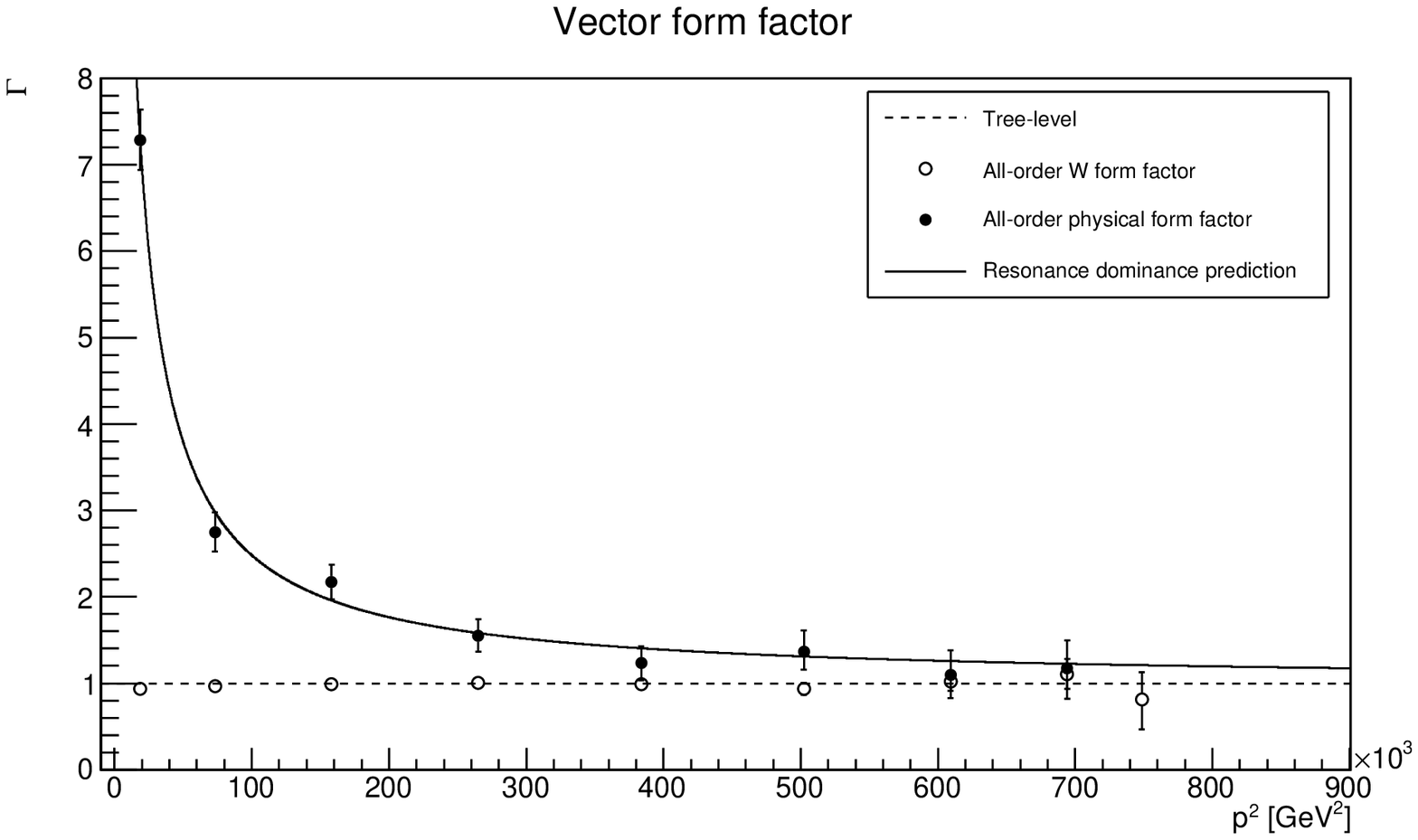}
\caption{\label{vertex}The physical vector boson form factor compared to the gauge dependent $W$/$Z$ form factor from \cite{Maas:2018ska} using lattice calculations at space-like momenta. The resonance dominance prediction is $a+bm_V^2/(p^2-m_V^2)$. In the conventions of \cite{Maas:2017wzi} the parameters are gauge coupling $g=1.15$, 4-Higgs coupling $\lambda=1.33$, and Higgs vacuum expectation value $v=69.6$ GeV.}
\end{figure}

One observable quantity are form factors. For the physical vector bosons, which reduce at leading order in the vacuum expectation value to the gauge-dependent $W^\pm$ and $Z$ bosons \cite{Frohlich:1980gj,Frohlich:1981yi,Maas:2017wzi}, one of the weak form factors has been investigated using lattice methods in \cite{Maas:2018ska}, and compared to the leading order in the vacuum expectation value. The result is shown in figure \ref{vertex}. At leading order in the vacuum expectation value FMS-augmented perturbation theory predicts the physical form factor to coincide with the gauge-dependent one. Just like for the propagator in figure \ref{pt}, this happens at large momenta.

That this happens at large momenta is understandable from physics intuitively, as here the substructure is resolved. This is quite similar to hadrons. At small energies, however, the composite state as a whole is probed. At weak coupling, the form factor should then be dominated by a single time-like pole \cite{Pacetti:2015iqa}. This is indeed the case, and yields a size parameter for the composite state of roughly $2/m_W$, relatively independent of the values of the couplings \cite{Maas:2018ska}. Measuring this form factor should be possible experimentally, though may require a lepton collider.

The second observable quantity are (differential) cross sections themselves in suitable processes. Just like for the Higgs, also weakly-charged fermions are not gauge-invariant. They therefore need to be also described by suitable composite operators \cite{Maas:2017wzi}, no matter whether they are leptons \cite{Frohlich:1980gj,Frohlich:1981yi} or open-flavor hadrons \cite{Egger:2017tkd}. Especially, this is also true for the proton. Similar as in \pref{higgs}, the leading contribution is again the ordinary proton, but subleading contributions have a Higgs component. This acts like an additional valence Higgs in the proton, and should be detectable in proton-proton collisions at sufficiently high energies.

Unfortunately, the proton structure is very complicated, and usually only effectively described with PDFs. Thus, a direct calculation using perturbation theory is not possible. However, following \cite{Egger:2017tkd}, also the valence Higgs could be captured by an additional PDF \cite{Fernbach:unpublished}.

To test this idea, an ad hoc Higgs PDF can be introduced. This can technically be realized using Herwig 7 event generator \cite{Bahr:2008pv,Bellm:2015jjp,Bellm:2017bvx} using the Matchbox module \cite{Platzer:2011bc} and the angular ordered shower \cite{Gieseke:2003rz}, with matrix elements provided by a combination of MadGraph5\_aMCatNLO \cite{Alwall:2014hca} and ColorFull \cite{Sjodahl:2014opa}. For details of this computation see \cite{Fernbach:unpublished}. The basic idea is to simply add an additional Higgs PDF, and in a first step only rescale the contribution of all partons. While a complete refit would certainly be preferable, this is at the moment too demanding in terms of resources. As a consequence, the valence Higgs contribution can be captured by the form of its PDF and a single parameter, which parametrizes the Higgs content of the proton.

\begin{figure}
\includegraphics[width=\linewidth]{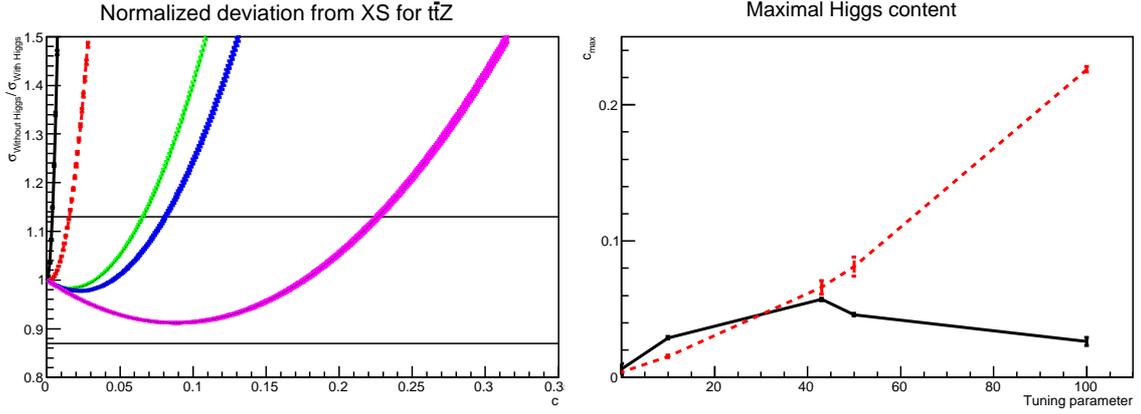}
\caption{\label{pdf}(PRELIMINARY RESULT, see \cite{Fernbach:unpublished} for the final result) The left-hand side shows the deviation of the cross section with valence Higgs contribution normalized to the one without for the process $pp\to\bar{t}tZ$ as a function of the Higgs 'content' of the proton for five values of the tuning parameter $c_1$ of a parameterized Higgs PDF $(1-x)exp(-c_1x^2)/x$ . The band-widths are statistical errors from Herwig. The horizontal bars indicate the size of the experimental error on the cross section of 13\% \cite{Sirunyan:2017uzs,Defranchis:2019mvt}. The right-hand plot gives the maximum Higgs content as a function of tuning parameter for this process (dashed) and the process $pp\to\bar{t}t$ (full line).}
\end{figure}

Processes which involve strong couplings to the Higgs are most suitable to test this contribution. Hence, processes involving tops and weak gauge bosons are obvious choices. A preliminary result is shown in figure \ref{pdf}. Final results can be found in \cite{Fernbach:unpublished}. Here, the impact of differently shaped valence Higgs PDFs with some Higgs fraction of the proton on the cross section for the processes $pp\to\bar{t}t$ and $pp\to\bar{t}tZ$ are shown. For a small enough Higgs content the effect is always compatible within errors with the experimental results. The maximum content depends on the shape of the PDF. Given the current expirmental precision \cite{Sirunyan:2017uzs,Defranchis:2019mvt}, a Higgs content of a few percent is still compatible with the data. A more elaborate analysis including several differential cross sections and detector uncertainties, however, reduces this fraction substantially \cite{Fernbach:unpublished}. Still, given the prospects for the high-lumi LHC and/or future machines, this can be a detectable effect. However, a full PDF refit including the valence Higgs PDF is desirable, to go beyond this exploratory study.

\section{Conclusions}

The field-theoretical necessity of gauge invariance at the non-perturbative level demands a slight augmentation of the Feynman rules using the FMS mechanism \cite{Frohlich:1980gj,Frohlich:1981yi,Maas:2017wzi}. This effect has been confirmed in lattice calculations \cite{Maas:2017wzi} and leads to slight deviations from results of not augmented perturbation theory \cite{Egger:2017tkd,Maas:2018ska,Fernbach:unpublished,Sondenheimer:unpublished}, which are in principle measurable. Herein a number of quantities have been outlined, for which first estimates of the size of the effect are already available. Many more are straightforward to obtain.

However, the real impact of this effect becomes visible beyond the standard model, in which various structural aspects reduces the impact \cite{Maas:2017wzi}. There, depending on the theory, qualitative differences are possible \cite{Maas:2015gma,Maas:2017xzh,Sondenheimer:gut}, as is also confirmed in lattice calculations \cite{Maas:2018xxu}. It is possibly crucial to take this into account in model-dependent searches for new physics.

\bibliographystyle{bibstyle}
\bibliography{bib}

\end{document}